%% file: main.tex
\documentclass[acmsmall]{acmart}
\usepackage{tikz}
\usepackage{mathtools, cuted}
\usepackage{lipani_math}
\usepackage{svg}
\usepackage{comment}
\usepackage{xcolor}
\usepackage{multirow}
\usepackage{enumitem}
\usepackage{fontawesome}
\usepackage{graphicx}
\usepackage{supertabular}
\usepackage{caption}
\usepackage{subcaption}
\usepackage{amsmath}
\usepackage{bm}
\usepackage{enumitem}

\makeatletter
\def\runningfoot{\def\@runningfoot{}}
\def\firstfoot{\def\@firstfoot{}}
\makeatother

\newcommand{\sizecirle}{0.8ex}
\newcommand{\rc}{\tikz\fill[blue] (0,0) circle (\sizecirle);}
\newcommand{\bc}{\tikz\fill[red] (0,0) circle (\sizecirle);}
\newcommand{\tc}{%
\begin{tikzpicture}
\fill[red] (0,0) circle (\sizecirle); 
\fill[blue] (0,0) -- (90:-\sizecirle) arc (-90:90:\sizecirle) -- cycle;
\fill[white] (0,0) circle (0.5ex); 
\end{tikzpicture}}

\newcommand{\lps}{\text{\faThumbsOUp}}

\newcommand{\las}{\text{\raisebox{-2pt}{\faThumbsODown}}}

\newcommand{\llp}{\,\text{\bc}}
\newcommand{\lcp}{\,\text{\rc}}
\newcommand{\lbp}{\,\text{\tc}}

\usepackage{breqn}
\begin{document}

\title[Political Bias Detection]{Quantifying Political Bias in News Articles}

\author{Gizem Gezici}
\affiliation{%
  \institution{Huawei Turkey R\&D Center}
  \city{Istanbul}
  \country{Turkey}}
\email{gizem.gezici@huawei.com}

\renewcommand{\shortauthors}{Gezici, et al.}

\begin{abstract}
Search bias analysis is getting more attention in recent years since search results could affect  In this work, we aim to establish an automated model for evaluating ideological bias in online news articles. The dataset is composed of news articles in search results as well as the newspaper articles. The current automated model results show that model capability is not sufficient to be exploited for annotating the documents automatically, thereby computing bias in search results.
\end{abstract}


\maketitle
\input{body.tex}

\bibliographystyle{ACM-Reference-Format}
\bibliography{references}
\end{document}

%% file: body.tex
\section{Introduction}
\label{sec:introduction}
\input{introduction}


\section{Background \& Related Work}
\label{sec:related_work}
\input{related_work.tex}

\section{Political Bias Evaluation Framework}
\label{sec:bias_evaluation_measures}
\input{political_bias_evaluation_framework.tex}

\section{Experimental Setup}
\label{sec:experiments}
\input{experiments.tex}

%% file: introduction.tex
Search engines are ubiquitous. As reported by SmartSights~\cite{SmartSights}, in 2017 46.8\% of the world population accessed the internet and by 2021, the number is expected to reach 53.7\%. Currently, on average 3.5 billion Google searches are done per day~\cite{InternetLiveStats}. These statistics advocate that search systems are ``gatekeepers to the Web'' for many people~\cite{diaz2008through}. 
As information seekers search the Web more, they are also more influenced by Search Engine Result Pages (SERPs) and their influence -negative included- potentially become visible in a wide range of areas. However, as with all software-based systems, search platforms do not lack of human influence, thereby they may suffer from embedded bias, i.e., corpus or algorithmic bias. Experimental studies suggest that particular information types or sources might be retrieved more or less, or might not be well represented~\cite{culpepper2018report}.
For instance, during the elections, it is known that people issue repeated queries on the Web about political candidates and events such as ``democratic debate'', ``Donald Trump'' and ``climate change''~\cite{kulshrestha2018search}.
\citet{epstein2015search} claim that SERPs returned in response to these queries may influence the voting decisions of the users and report that \textit{manipulated} search rankings can change the voting preferences of undecided individuals at least by 20\%. As individuals rely to a greater extent on the SERPs for their decision making, there is a thriving demand for these systems explainable.


%
Empirical evidence has shown that individuals trust more sources ranked at higher positions in SERPs, but the ranking criteria may rather depend more on user satisfaction than the factual information, which jeopardizes the phenomenon of providing reliable information in exchange for satisfying users~\cite{culpepper2018report}.
In spite of the given research findings, the majority of online users tend to believe that search engines provide \emph{neutral} results, i.e., serving only as facilitators in accessing information on the Web due to their automated operations~\cite{goldman2008search}. However, this romanticised view of search platforms does not reflect reality and there seems to be a growing skepticism related to objectivity and credibility of these platforms. 
To illustrate that, a recent dispute between the U.S.~President Donald Trump and Google can be given, where Mr.~Trump accused Google of presenting only negative news about him when his name is searched. Google refuted this claim by saying that: ``When users type queries into the Google Search bar, our goal is to make sure they receive the most relevant answers in a matter of seconds'' and ``Search is not used to set a political agenda and we don't bias our results toward any political ideology''\footnote{https://www.reuters.com/article/us-usa-trump-tech-alphabet/google-responds-to-trump-says-no-political-motive-in-search-results-idUSKCN1LD1QP}. In this work, we hope to shed some light on that debate, by not specifically examining queries regarding Donald Trump but by fulfilling an in depth analysis of retrieved search answers to a broad set of queries related to controversial topics based on concrete evaluation measures.

%

\emph{Bias} implies undue emphasis. For a retrieval system, it can be defined as the balance and representativeness of Web documents retrieved from a database for a set of queries~\cite{mowshowitz2002assessing}.
When a user issues a query to a search engine, 
documents from different sources are collected, ranked, and presented to the user. Assume that a user searches for \emph{2016 presidential election} and the top-n ranked results are displayed. In such a search scenario, the retrieved results may exaggerate or downplay particular perspectives and thereby 
provide an unbalanced picture of
the given query as claimed by Mr.~Trump, though without any scientific support.
Hence, 
the potential \emph{undue inclusion or exclusion} of specific perspectives in the retrieved results lead to bias~\cite{mowshowitz2002assessing}. Note that the existence of bias is different from relevance. In the presented scenario, even though the retrieved list of documents may all be judged as relevant with respect to the given query, if the selection of the documents is skewed or slanted, i.e., emphasizing one perspective over another, then the corresponding search engine is \emph{biased} due to an imbalanced representation of the perspectives towards the query's topic. 
Bias is especially important if the query topic is \emph{controversial} having opposing perspectives as described in the given scenario. The bias in SERPs can be used by search engines to inform their users about the bias by making themselves more accountable which is one of the crucial attributes that a retrieval system should possess~\cite{culpepper2018report}.

In this work, we focus on the SERPs coming from the news sources and investigate two major search engines (Bing and Google) in terms of political bias. Our analysis has mainly three sides where we evaluate the political bias of the search engines separately, then make a comparison among them as well as track the source of this bias in the SERPs. The bias may come from the data, which may contain biases (input bias) or the search algorithm, which contains sophisticated features (algorithmic bias). In the scope of this work, we concentrate on the input bias that is intrinsically embedded in the data itself. 

In short, we aim to answer the following research questions: 
\begin{description}
    \item[RQ1:] On a conservative-to-liberal scale, do search engines return \emph{politically biased} SERPs in response to queries related to controversial topics? 
    \item[RQ2:] Are search engines \emph{significantly different} from each other towards controversial topics?    
    \item[RQ3:] Does the \emph{source of bias} come from the input data?
\end{description}

We address these research questions for \textit{controversial} topics representing a broad range of issues in SERPs of Google and Bing through content analysis, i.e. analysing the textual content of the retrieved documents. We focus on \emph{content bias} and describe our SERP politically content bias quantification framework in which we propose three different measures of bias based on common Information Retrieval (IR) \emph{utility-based} evaluation measures: Precision at cut-off (P@n), Rank Biased Precision (RBP), and Discounted Cumulative Gain at cut-off (DCG@n). 
While the first measure quantifies bias considering only a weak ranking criterion, i.e.~the first $n$ returned documents as in SERPs, the other two measures incorporate stronger ranking bias.

In order to answer RQ1, we measure the degree of deviation of the ranked SERPs from an \emph{ideal} distribution where different political perspectives are \emph{equally} likely to appear~\cite{mowshowitz2002assessing}. 
To detect political bias which results from the imbalanced representation of politically different points of view, we label the documents' political perspectives with classification and use these labels for bias evaluation.
To address RQ2, we compare the political bias in the SERPs of the two search engines to see if they show similar bias for the corresponding controversial topics.
To answer RQ3, we measure the political bias in a similar manner as we do to answer the RQ1. The only difference is that this time we measure the bias in the whole corpus, i.e., all retrieved SERPs from Google and Bing, instead of the partial of it, as top $n$ documents. Then, we compare the bias of the whole corpus with the partial bias that we compute for the first part of our analysis to check if they are consistent, or not.


Our main contributions in this work can be summarized as follows:

\begin{enumerate}
\item We propose a \emph{novel generalizable search bias evaluation framework} to measure the bias in the SERPs.
\item Our fairness-aware set of measures are explainable utility-based IR measures which take into account of relevance as well, while evaluating bias in ranked results.
\item We apply our framework to \emph{compare the relative bias} in the SERPs content of two search engines for controversial queries. To the best of our knowledge, this is one of the first works to compare the two search engines in terms of \emph{content bias}.
\item We utilize the framework to \emph{track the source of bias} in SERPs, i.e., checking if the existing bias comes from the corpus or not.
\end{enumerate}


Lastly, a preliminary version of the present framework has been published earlier~\cite{Gezici2021EvaluationResults}. This work expands and surpasses the previous endeavor mainly in the following points: (i) We used a deep learning framework that is widely-used in NLP tasks for classification, i.e., to predict the political perspectives of the retrieved documents from the two popular search engines, instead of obtaining these labels via crowdsourcing. (ii) We also crawled a different dataset of articles from the newspapers' websites to fine-tune the deep learning model specifically for the news articles, thereby achieve a better classification model in perspective detection. (iii) Finally, we included a new investigation on the source of bias in our experiments through examining if the measured bias in the top-n documents comes from the data.

The remainder of the paper is structured as follows. In Section \ref{sec:related_work} we present the related work. The search bias evaluation framework is developed in Section \ref{sec:bias_evaluation_measures}. 
In Section \ref{sec:experiments} we detail the experimental setup, and present the results.

%% file: related_work.tex
Studies indicate that online IR platforms have influenced public by causing an increase in polarization; for instance it has been shown that social media accounts with non-Western names are more likely to be indicated as fraudulent since most probably the system has been trained on Western names. Therefore, proposing new metrics can be useful in evaluating and understanding such ethical issues in IR systems~\cite{culpepper2018report}. Regarding this assertion, there have been a growing interest in quantifying bias in the recent years 


%% file: political_bias_evaluation_framework.tex
2In this section, to answer the research questions asserted in the introduction, we propose our political bias evaluation framework. The first question focuses on detecting political bias (if exists) in SERPs and the second one examines if the search engines show the same level of bias. On the other hand, the third question investigates if the source of bias comes from the data. In this framework, we initially identify political perspectives of the news articles automatically via classification. Then, we present the measures of bias and the proposed protocol to identify political bias and investigate the source of bias in SERPs.

\subsection{Preliminaries}
Our first goal is to detect political bias with respect to the distribution of political perspectives expressed in the contents of the SERPs.  
Let $\set{S}$ be the set of search engines and $\set{Q}$ be the set of queries about controversial topics. 
When a query $q \in \set{Q}$ is issued to a search engine $s \in \set{S}$, the search engine $s$ returns a SERP $r$. 
We define the political perspective of the $i$-th retrieved document $r_i$ with respect to $q$ as $j(r_i)$. A political perspective can have the following values:
\emph{conservative}, 
\emph{liberal}, 
\emph{both or neither}. 

A document political perspective can be:
\begin{description}
\item[conservative ($\lcp\,$)] when the document content is in favour of the conservative political agenda.
\item[liberal ($\llp\,$)] when the document content is in favour of the liberal political agenda.
\item[both or neither ($\lbp\,$)] when the document content is in favour of both or neither perspectives.
\end{description}

For our analyses, we deliberately use \emph{controversial} topics such as abortion, medical marijuana, gay marriage, and Cuba embargo which contain opposing viewpoints since complicated concepts concerning the identity, religion, or political leaning are the actual points where search engines are more likely to provide biased results~\cite{noble2018algorithms}. Our second goal is related to the first one as we compare the search engines in terms of the political bias that they show in their SERPs.

Our third goal is to find the source of political bias by checking if the bias comes from the data (input bias). We do this by measuring the bias also in the whole corpus of the SERPs in addition to the partial dataset, i.e. top 10 or 100 documents of the corpus, and comparing the corpus bias (if exists) with the partial one to see if they are consistent or not.

\subsection{Measures of Bias}
\label{subsec:biasmeasures}
On the basis of the previously definition of bias given in Section \ref{sec:introduction}, it can be quantified by measuring the degree of deviation of the distribution of documents from the \textit{ideal} one. To give a generic definition of an ideal list in the field of bias evaluation poses problems; but in the scope of this work, we can assess the existence of political bias in a ranked list retrieved by a search engine if the presented information \emph{significantly deviates} from true likelihoods~\cite{white2013beliefs}. Using this definition reversely, we can adopt the assumption that the \textit{ideal} list is the one that minimizes the difference between two opposing political perspectives, which we indicate here as $\lcp\,$ and $\llp\,$.

\subsubsection{Bias Score of a SERP} 
Formally, we measure the political bias in a SERP $r$ as follows:
\begin{equation}\label{eq:bias}
    \beta_f(r) = f_{\lcp\,}(r) - f_{\llp\,}(r),
\end{equation}
where $f$ is a function that measures the likelihood of $r$ in satisfying the information need of the user about the view $\lcp\,$ and the view $\llp\,$.
\subsubsection{Mean Bias (MB)}
Before defining $f$, from Eq.~\eqref{eq:bias}, we define the mean bias of a search engine $s$ as:
\begin{equation}\label{eq:expected_bias}
    \text{MB}_f(s, \sQ) = 
    \frac{1}{|\sQ|}
    \sum_{q \in \sQ} \beta_f(s(q)).
\end{equation}
An unbiased search engine would produce a mean bias of $0$. 
A limitation of MB is that if a search engine is biased towards the $\llp\,$ perspective on one topic and bias towards the $\lcp\,$ perspective on another topic, these two contributions will cancel each other out. 
In order to avoid this limitation we also define another metric.
\subsubsection{Mean Absolute Bias (MAB)}
The mean absolute bias, which consists in taking the absolute value of the bias for each $r$. Formally, this is defined as following:
\begin{equation}\label{eq:expected_bias}
    \text{MAB}_f(s, \sQ) = 
    \frac{1}{|\sQ|}
    \sum_{q \in \sQ} |\beta_f(s(q))|.
\end{equation}
An unbiased search engine produces a mean absolute bias of $0$. Although this measure solves the limitation of MB, MAB says nothing about towards which perspective the search engine is biased, making these two measures of bias complementary.

\subsubsection{Retrieval Evaluation Measures}
In IR the likelihood of $r$ in satisfying the information need of users is measured via retrieval evaluation measures. Among these measures we selected 3 \emph{utility-based} evaluation measures. This class of evaluation measures quantify $r$ in terms of its worth to the user and are normally computed as a sum of the information gain summed over the relevant documents retrieved by $r$. 

The 3 IR evaluation measures used in the following experiments are: 
P$@n$, RBP, and DCG$@n$.

\begin{description}
\item[P$@n$:]
P$@n$ for the $\lps$ view is formalized in~\cite{Gezici2021EvaluationResults}. However, differently from the previous definition of $j$ where the only possible outcomes are $g_1$ and $g_2$, here $j$ can return any of the label associated to a political perspective ($\llp\,$, $\lcp\,$, and $\lbp\,$). Hence, only conservative and liberal documents, that is relevant to the topic, are taken into account, since $j(r_i)$ returns \emph{both or neither} 
when otherwise.

Replacing P$@n$ in Eq.~\eqref{eq:bias} we obtain the first measure of bias:
\begin{equation}\label{eq:bias_pc}
    \beta_{\text{P}@n}(r) = 
    \frac{1}{n}\sum_{i=1}^n 
    \left([j(r_i) = \lps] - [j(r_i) = \las]\right).
\end{equation}
The main limitation of this measure of bias is that it has a weak concept of ranking, 
i.e.~the first $n$ documents contribute equally to the bias score. The next two evaluation measures overcome this issue by defining discount functions.

\item[RBP:]
RBP weights every document based on the coefficients of a normalized geometric series with value $p \in ]0,1[$, where $p$ is a parameter of RBP. Similarly to what is done for P$@n$, we reformulate RBP to measure bias as follows:
\begin{equation}\label{eq:rbpc}
    \text{RBP}_{\lps} = 
    (1-p) 
    \sum_{i=1} p^{i-1}[j(r_i) = \lps].
\end{equation}

Substituting Eq.~\eqref{eq:rbpc} to Eq.~\eqref{eq:bias} we obtain:
\begin{equation}\label{eq:bias_rbpc}
    \beta_{\text{RBP}}(r) = (1-p) \sum_{i=1} p^{i-1} \left(
    [j(r_i) = \lps] - [j(r_i) = \las]
    \right).
\end{equation}

\item[DCG$@n$]
DCG$@n$, instead, weights each document based on a logarithmic discount function. Similarity to what is done for P$@n$ and RBP, we reformulate DCG$@n$ to measure bias as follows:
\begin{equation}\label{eq:dcgc}
    \text{DCG}_{\lps}@n = \sum_{i=1}^{n} 
    \frac{1}{\log(i + 1)}[j(r_i) = \lps].
\end{equation}

Substituting Eq.~\eqref{eq:dcgc} to Eq.~\eqref{eq:bias} we obtain:
\begin{equation}\label{eq:bias_dcg}
    \beta_{\text{DCG}@n}(r) = \sum_{i=1}^{n}
    \frac{1}{\log(i + 1)}\left(
    [j(r_i) = \lps] - [j(r_i) = \las]
    \right)
\end{equation}

\end{description}

Since we are evaluating web-users, for P$@n$ and DCG$@n$ we set $n=10$ and for RBP we set $p = 0.8$.

This last formulation (Eq.~\label{eq:bias_dcg}), although it looks similar to the rND measure, it does not suffer from the first four limitations introduced in Section \ref{sec:related_work}. 
In particular all these presented measures of bias:
1) do not focus on one group;
2) use a binary score associated to the document stance or political perspective, similar to the way these measures are used in IR when considering relevance; also like in IR  
3) can be computed at each rank; 
4) exclude non-relevant documents from the measurement of bias, and; 
5) provide various user models associated to the 3 IR evaluation measures: P@n, DCG@n, and RBP.

\begin{figure*}[!t]
  \centering  
  \includegraphics[width=\textwidth]{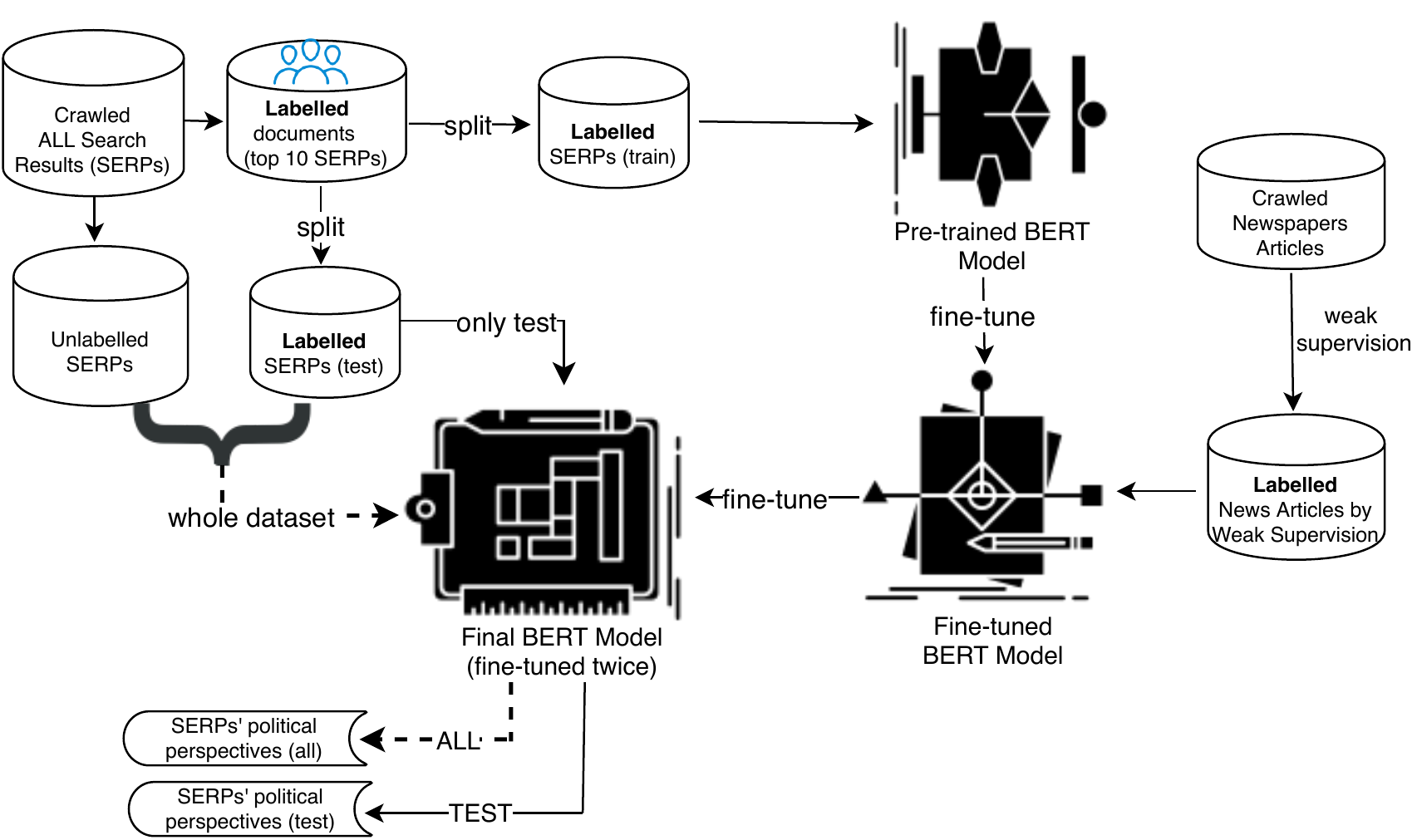}
  \caption{Flow-chart of the quantifying political bias procedure.}
  \label{fig:flowchart}
\end{figure*}

\subsection{Quantifying Political Bias}
\label{subsec:protocol}
Using the previously defined measures of bias in Section \ref{subsec:biasmeasures} we quantify the political bias of the two search engines, Bing and Google. Then we compare them in terms of the measured 
bias. Now, we describe each step of the proposed mechanism used to quantify political bias in SERPs.

\begin{description}
\item[Dataset Construction.]
After having crawled all the SERPs for both search engines and all queries $\set{Q}$, for each returned document we obtain the political perspective of the document with respect to the corresponding query in the top 10 documents and the whole corpus. Both is done automatically via classification since we already have a dataset labelled by crowdsourcing for training.
\item[Bias Evaluation.]
We compute the bias measures of every SERP with all three IR-based measures of bias: P$@n$, RBP, and DCG$@n$ for the first 10 documents and only P$@n$ for the whole corpus, because the other measures of bias do not provide meaningful results due to the rank information used in these measures. We then aggregate the results using the two measures of bias, MB and MAB.
\item[Statistical Analysis.]
To identify whether the bias measured is not rooted from randomness, for the bias measure MAB, we compute a one-sample t-test: 
the null hypothesis is that no difference exists and that the true mean is equal to zero.
If this hypothesis is rejected, this means that there is a significant difference; we claim that the evaluated search engine is biased.
Then, we make a comparison in the bias difference measured across the two search engines using a two-tailed paired t-test: 
the null hypothesis is that the difference between the two true means is equal to zero.
If this hypothesis is rejected, this means that there is a significant difference so we claim that there is a difference in bias between the two search engines.

\end{description}
\

%% file: experiments.tex
\begin{figure*}[!t]
    \centering
    {\includegraphics[height=0.6\textwidth,trim={0.5cm 0.1cm 0.8cm 0.8cm},clip]{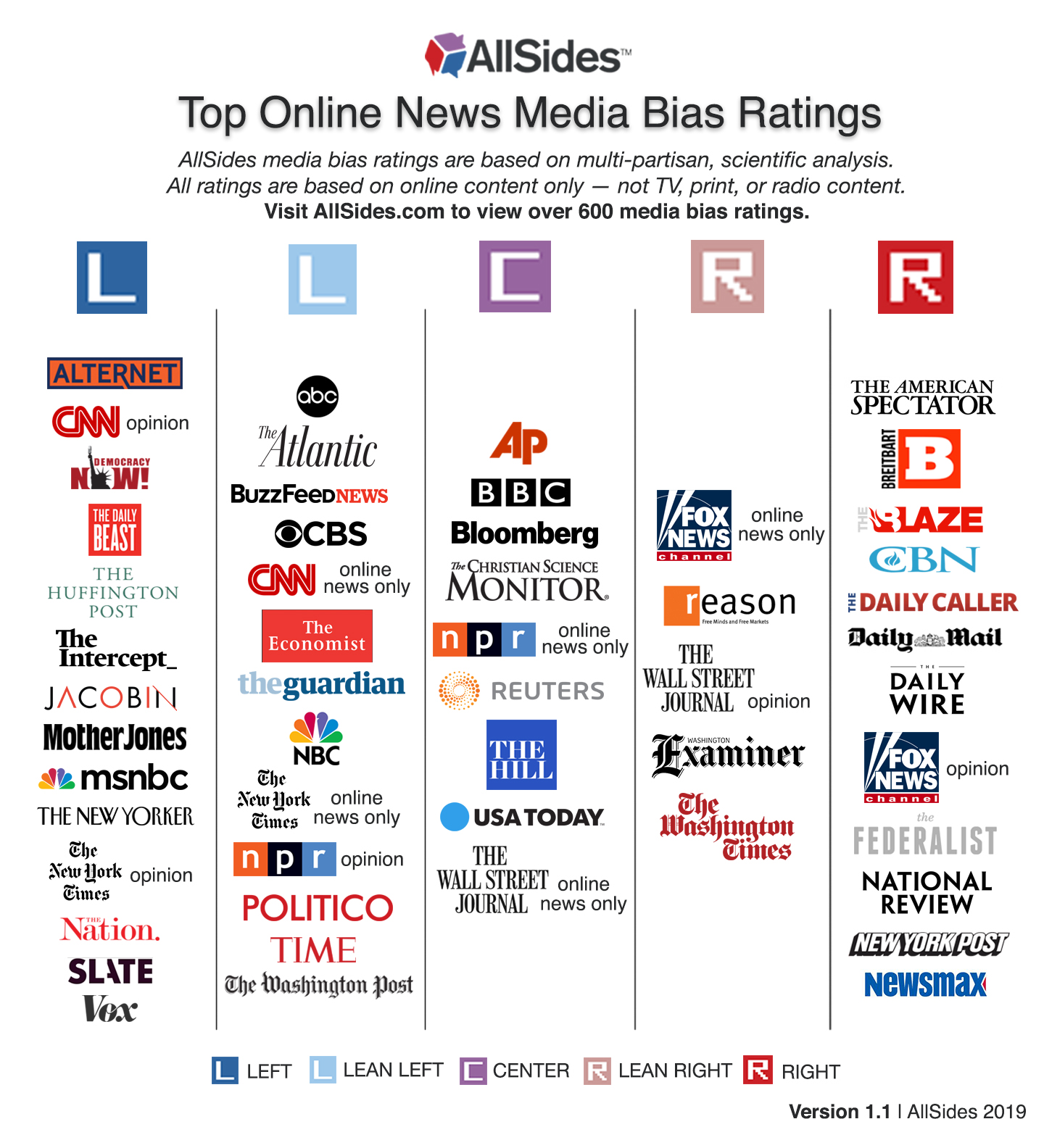}}
    \caption{Top Online News Media Bias Ratings of AllSides}
    \label{fig:mediabias}
\end{figure*}

\subsection{Dataset}
\label{subsec:dataset}
In this work, we used two datasets for political bias evaluation. The first dataset contains SERPs crawled from two search engines, Google and Bing. The second dataset, on the other hand is composed of the articles published in the known newspapers. 

\begin{description}
\item[News Articles in SERPs.] We obtained the controversial queries issued for searching from ProCong.org~\cite{ProCon} and applied some filtering steps on the initial query set. After filtering, the final query set size became 57. We submitted each query in the final query set to the US News search engines of Google and Bing using a US proxy. Then, we extracted the whole corpus returned by both engines in response to all the queries in the set. Note that the data collection process was done in a controlled environment such that the queries are sent to the search engines at the same time. For more details about the selection of the queries and crawling the SERPs, please refer to our previous work~\cite{Gezici2021EvaluationResults}. After having crawled all the SERPs returned from both engines and extracted their contents, we annotated the top 10 documents. We obtained the stance label of each document with respect to the queries via crowdsourcing. To label the political perspective of queries, we also used crowdsourcing. To obtain the political perspective of documents, we transformed the stance labels into political perspectives based on the perspective of their corresponding queries. The details about our crowdsourcing campaigns as well as the transformation process can also be found in~\cite{Gezici2021EvaluationResults}. 
\item[News Articles in Newspapers.] To enrich the pre-trained BERT model on perspective detection task even more, we crawled the article contents of the selected newspapers. To choose the newspapers to be crawled, we prepared a list by utilizing the media bias chart of AllSides.com \citet{AllSides}. AllSides.com is a US-based online website that shares information and ideas from all sides of the political spectrum with the purpose of fighting filter bubbles and polarization. The organization has researched since 2012, uses multiple methodologies to rate media bias and even achieved a patent on this area. Furthermore, AllSides only focuses on American sources and its media bias rating scale is based on American politics for the online news articles. For all these reasons, we chose the media bias chart of AllSides to determine the political perspective of newspapers which is displayed in Figure~\ref{fig:mediabias}. Subsequently, we crawled the textual contents of the news articles of the corresponding newspapers in the chart by using a Python library called feedparser~\footnote{https://pythonhosted.org/feedparser/} that parses feeds in all known formats such as Atom, RSS, and RDF. Then, we labelled the crawled news articles by weak supervision such that labelling the articles of left/lean-left newspaper as liberal, right/lean-right as conservative, and center as neutral. We further used this newspaper dataset to fine-tune the BERT model after the first fine-tuning done with the training data by learning the topic keywords and semantic relations in the articles, thereby detecting the political perspectives more accurately. 

\end{description}

\subsection{Classification}
\label{subsec:classification}
To answer the first research question, we used the political perspective labels of the top 10 documents in the first dataset previously obtained via crowdsourcing. In the scope of this work, we used the labelled dataset to obtain those perspectives automatically. For automatic perspective detection, we used pre-trained Bidirectional Encoder Representations from Transformers (BERT) model which is proposed by Google that presents state-of-the-art results in a wide variety of NLP tasks~\cite{devlin2018bert}. BERT's key difference from other deep models is having an attention mechanism called Transformer that applies bidirectional training as opposed to directional models to language modelling. In this way, BERT learns contextual relations between words from all of the surroundings (left and right) of a given word. 

We splitted the labelled dataset into train and test sets and fine-tuned the pre-trained BERT model  on the train test. Moreover, for making the BERT learn semantic clues better in the SERPs, in this work we crawled the newspapers' contents as the second dataset which is similar to those news articles in the SERPs, labelled with weak supervision and used this dataset to fine-tune the BERT model again. To secondly fine-tune the already fine-tuned BERT model, we used weak supervision labels of the news articles as depicted in Figure~\ref{fig:flowchart}. As the lower-quality labels were being updated by the existing BERT model, the model was expected to learn linguistic clues such as semantic relations and important keywords which belong to a specific category (conservative, liberal, or neutral) simultaneously. The second question is highly related to the first one, thus we directly used the results of the first question which showed the levels of political bias in the SERPs of Bing and Google, then made a comparison among them by using the statistical analyis described in Section~\ref{subsec:protocol}. 

To address the third question, we used the whole corpus in the first dataset to investigate if the input bias exists. To track the source of bias, we made use of the final (fine-tuned twice) BERT model to label all the SERPs in the first dataset and check if the bias in the top 10 documents is consistent with the whole corpus.

\subsection{Results}
\label{subsec:results}

\begin{table}[]
\centering
\caption{Perspective dataset}
\begin{tabular}{p{1.2cm}|p{1.2cm}|p{1.2cm}||c|c}
\toprule
\multirow{2}{*}{\textbf{query}} & 
\multicolumn{2}{c||}{\textbf{document}} & 
\multicolumn{2}{c}{\textbf{Weak Supervision}} \\ 
\cline{2-5}
& 
\textbf{title} & \textbf{content} & \textbf{Without} & \multicolumn{1}{c}{\textbf{With}} \\ 
\hline
 & \checkmark &  &  0.647 & 0.692 \\ \hline
 &  \checkmark &  \checkmark &  0.692 & 0.692 \\ \hline
 \checkmark &  \checkmark &  &  0.676 & 0.707 \\ \hline
 \checkmark &  \checkmark &  \checkmark  & 0.707 & 0.723 \\
 \bottomrule
\end{tabular}
\end{table}

Exp. 1: Fine-tuning BERT \textbf{on only perspective data} using the following inputs:

\begin{enumerate}[label=\alph*)]
    \item only document content with title
    \item concatenating query and document content with title
    \item only document title
    \item concatenating query and document title
\end{enumerate}

Exp. 2: Fine-tuning BERT \textbf{on only news data} using the following inputs:

\begin{enumerate}[label=\alph*)]
    \item only document content with title
    \item only document title
\end{enumerate}